\documentclass[showpacs,preprintnumbers,amsmath,amssymb]{revtex4}
\usepackage{graphicx}
\usepackage{bm}
\usepackage{amsfonts}
\usepackage{indentfirst}
\usepackage[small]{caption2}
\usepackage{longtable}

\begin{document}


\title{On evolution of the purity of two nonidentical two-level atoms interecting with one-mode coherent field}
\author{E.K. Bashkirov}
 \altaffiliation[Electronic address:]{bash@ssu.samara.ru}
\affiliation{%
Department of General and Theoretical Physics, Samara State University, Acad. Pavlov Str.1 , 443011 Samara, Russia
\\
}%

\date{\today}

 \begin{abstract}
In this brief report the resonant interaction of two nonidentical two-level atoms with one mode of the electromagnetic field has been considered.
 The pure-state evolution of the atomic states for field initially in the coherent state and atoms in the ground state
 has been investigated. It has been shown that for intermediate values of the relative differences of two coupling
  constants the  atoms as well as the  field are returned most closely to a pure state at the revival time.
 The possibility of the maximally entangled states at the beginning of the collapse time has been discussed.

\end{abstract}



\maketitle

The Jaynes-Cumming model (JCM)\cite{Shore} is the simplest and most effective model to describe a two-level atom interacting with electromagnetic field. The interaction of an atom with quantized field leads to entanglement of these two subsystem so that the total state vector cannot be written precisely as the product of time-dependent
 atomic and field component vectors. Quantum entanglement has attracted much attention in recent years due to its
 potential applications in quantum communication, information processing and quantum computing \cite{Yamamoto},\cite{Di}.
  Entanglement is well understood for pure states of bipartite systems such as JCM. In this case, entanglement
   is quantified by the entropy of the reduced atomic or field density matrix.
   The more sensitive measure of entanglement and correlations of subsystem is the von Neumann entropy of either of the subsystem defines as
   \cite{Knight}
   \begin{eqnarray}
S  = -Tr[\rho_{at(field)} \ln \rho_{at(field)}].\nonumber
\end{eqnarray}
   Alternatively, one can consider the quantity $Tr (\rho_{at})^2$ (or so-called linear atomic entropy S = 1 - $Tr (\rho_{at})^2$) as a measure of degree of the purity of the atomic
   states  and on the whole the entanglement between atomic and field subsystems \cite{Ban}. In this case, the atomic subsystem is said to be in the pure state if the condition
   $Tr (\rho_{at})^2 = 1$ is satisfied and in the mixed state if  $Tr (\rho_{at})^2 < 1$. The maximally entangled mixed state corresponds
   to  $Tr (\rho_{at})^2 = 1/2$ for one-atom JCM. The considered approach
   has been used to analyse the purity evolution for one-mode JCM \cite{Ban}, two-mode JCM \cite{Nas} and two-atom JCM
   with identical atoms \cite{H}. The purity of the field state or linear field entropy in Jaynes-Cummings like models has considered in \cite{Field}.
   As in the case with von Neumann entropy,   one can make the same principal conclusions about atomic
    purity and entanglement with analyzing the time behaviour of $Tr (\rho_{at})^2$ for JCM.
   The main purpose of this paper is to investigate the evolution of correlations or entanglement between two nonexcited
   nonidentical two-level atoms and mode of cavity field initially prepared in coherent state. In this model
   only the alternative
   approach allows us to consider the  analytical expression for linear entropy.

    Let us consider a system of two nonidentical two-level atoms interacting with a single-mode quantized electromagnetic
  field in a lossless resonant cavity via the one-photon-transition mechanism. The Hamiltonian for
  the considered system in the rotating wave approximation is
\begin{eqnarray}
H  =  \hbar \omega a^+ a + \sum\limits_{i=1}^2 \hbar \omega_0 R^z_i + \sum\limits_{i=1}^2 \hbar g_i (R^+_i a + R^-_i a^+), \label{1}
\end{eqnarray}
 where $a^+$ and $a$ are the creation and annihilation operators of photons of the cavity field, respectively,
 $R^+_f$ and $R^-_f$ are the raising and the lowering operators for the $i$th atom, $\omega$ and $\omega_0$ are
 the frequencies of the field mode and the atoms, $g_i$ is the coupling constant between the $i$th atom and the field.
 We assume the field to be at one-photon resonance with the atomic transition, i.e. $\omega_0 = \omega$ .

We denote by  $\mid + \rangle$ and $\mid - \rangle$ the excited and ground states of a single two-level atom and by
 $\mid n\rangle$ the Fock state of the electromagnetic field. The two-atom wave function can be expressed as
 a combination of state vectors of the form $\mid \it v_1, \it v_2 \rangle = \mid \it v_1\rangle \mid \it v_2 \rangle $,
 where $\it v_1, \it v_2 = +,-$.
   Let  the atoms are initially in the ground state $\mid -, -\rangle$ and the field is initially in a coherent state
   $\mid z \rangle$,
\begin{eqnarray}
\mid z \rangle = \sum\limits_{n=0}^{ \infty}\exp
 \left (-\frac{\mid z \mid^2}{2}\right )
 \frac{z^n}{\sqrt{n!}},\nonumber
 \end{eqnarray}
 where $ z = \mid z \mid
 e^{\imath \varphi}$ and $ \overline{n} = \mid z \mid^2$ is
  the initial mean photon number or dimensionless intensity of the  cavity field.

  The time-dependent wave function of the total system $\mid \Psi(t) \rangle$ obeys the ${\rm Schr\ddot{o}dinger}$ equation
\begin{eqnarray}
\imath \hbar  \mid \dot\Psi(t) \rangle = H \mid \Psi(t) \rangle.
\end{eqnarray} Using the Hamiltonian (1) the wave function in the interaction representation
  is found to be
 \begin{eqnarray}
 \mid \Psi(t) \rangle  = && \sum\limits_{n=0}^{ \infty} \exp
 \left (-\frac{\mid z \mid^2}{2}\right )
 \frac{z^n}{\sqrt{n!}}\times \nonumber\\
 &&\times[ C^{(n)}_1(t) \mid + , +; n-2 \rangle + C^{(n)}_2(t) \mid + , -; n-1 \rangle +
  + C^{(n)}_3(t) \mid - , +; n-1 \rangle +
   C^{(n)}_4(t) \mid - , -; n  \rangle ].
   \end{eqnarray}

For atoms initially prepared in their ground state we have the initial conditions for probability  coefficients
\begin{eqnarray}
C^{(n)}_4(0) =1,\quad C^{(n)}_1(0)=C^{n)}_2(0) = C^{(n)}_3(0) =0\quad (n=0,1,2,\ldots).
\end{eqnarray}

With the help of formulas (1)-(4) one can obtain  the time-dependent  probability  coefficients $C^n_i(t)$ in the form
\begin{eqnarray}
C^{(0)}_1(t) =  C^{(0)}_2(t)=C^{(0)}_3(t) = 0,\quad C^{(0)}_4(t) = 1; \nonumber \\
C^{(1)}_1(t) = 0,\quad C^{(1)}_2(t) = \frac {-\imath \sin(\sqrt{1+R^2}t)}{\sqrt{1+R^2}},\nonumber \\
C^{(1)}_3(t) = \frac {-\imath R\sin(\sqrt{1+R^2}t)}{\sqrt{1+R^2}}, \quad C^{(1)}_4(t) = \cos(\sqrt{1+R^2}t)
 \end{eqnarray}
and for $n\geq 2$
\begin{eqnarray}
 C^{(n)}_1(t) = \frac {2 R\sqrt{(n-1)n}}{\beta}[\cos(\lambda_+ t) - \cos(\lambda_- t)],\nonumber \\
 C^{(n)}_2(t) = \frac {-4 \imath R^2 (n-1)\sqrt{n}}{\beta}
 \left \{\frac{\lambda_+^2 + (1-R^2) n}{\lambda_+[\beta-(1+R^2)]} \sin(\lambda_+ t) -
\frac{\lambda_-^2 + (1-R^2) n}{\lambda_-[\beta+(1+R^2)]} \sin(\lambda_- t)
 \right \}, \nonumber \\
C^{(n)}_2(t) = \frac {-4 \imath R (n-1)\sqrt{n}}{\beta}
 \left \{\frac{\lambda_+^2 - (1-R^2) n}{\lambda_+[\beta-(1+R^2)]} \sin(\lambda_+ t) -
\frac{\lambda_-^2 - (1-R^2) n}{\lambda_-[\beta+(1+R^2)]} \sin(\lambda_- t)
 \right \}, \nonumber \\
 C^{(n)}_4(t) = \frac {8 R^2(n-1)n}{\beta}\left [\frac{\cos(\lambda_+ t)}{\beta-(1+R^2)} +
 \frac{\cos(\lambda_- t)}{\beta+(1+R^2)}\right ],
 \end{eqnarray}
 where
\begin{eqnarray}
\lambda_{\pm} = \sqrt{(1+R^2)(2n-1) \pm \beta}/\sqrt{2},\nonumber \\
 \beta =\sqrt{(2n-1)^2(1+R^2)^2 - 4 (n-1)n(1-R^2)^2}, \quad R=g_2/g_1 \nonumber .
\end{eqnarray}
  Note, that the explicit  form of coefficients $C^{(n)}_i$ for considered model   has been presented  firstly in
 \cite{Zubairy} but in these expressions some missprints take place.

Taking into account  formula  (3) one can write the quantity $Tr (\rho_{at})^2 $ in the following form
\begin{eqnarray}
 Tr (\rho_{at})^2 = &&\overline{n}^4 \left [\sum\limits_{n=0}^{\infty} \frac{ p_n}{(n+1)(n+2)} \mid C^{(n+2)}_1 \mid^2
 \right ]^2 + \nonumber \\
&&+\overline{n}^2 \sum\limits_{i=2}^3 \left [\sum\limits_{n=0}^{\infty} \frac{ p_n}{(n+1)} \mid C^{(n+1)}_i \mid^2
 \right ]^2 +  \left [\sum\limits_{n=0}^{\infty}  p_n \mid C^{(n)}_4 \mid^2
 \right ]^2 + \nonumber \\
 &&  +2 \overline{n}^3 \sum\limits_{i=2}^3 \left [ \left | \sum\limits_{n=0}^{\infty} \frac{ p_n}{(n+1)\sqrt{n+2}}
  C^{*(n+1)}_i C^{(n+2)}_1 \right |
 \right ]^2 + 2 \overline{n} \sum\limits_{i=2}^3 \left [ \left | \sum\limits_{n=0}^{\infty} \frac{ p_n}{\sqrt{n+1}}
  C^{*(n)}_4 C^{(n+1)}_i \right |
 \right ]^2 + \nonumber \\
&&+ 2 \overline{n}^2\left [ \left | \sum\limits_{n=0}^{\infty} \frac{ p_n}{\sqrt{(n+1)(n+2)}}
  C^{*(n)}_4 C^{(n+2)}_1 \right |
 \right ]^2 + 2 \overline{n}^2 \left [ \left | \sum\limits_{n=0}^{\infty} \frac{ p_n}{(n+1)}
  C^{*(n+1)}_3 C^{(n+1)}_2 \right |
 \right ]^2.
\end{eqnarray}

With  using  Eqs. (5)-(8) we can now proceed to calculate the behaviour of  $Tr (\rho_{at})^2$ for different values of the relative differences of two coupling constants $R = g_2/g_1$.

In Fig. 1 we have plotted the $Tr (\rho_{at})^2$ versus scaled time $g_1 t$ for coupling differences $R=0,\>\> 0.5 $ and $1$ when the field is initially in a coherent state with $\overline{n} = 50$. Obviously, that $R =0$ and $1$ correspond to a single two-level atom and
 two  identical two-level atoms respectively. To facilitate further analysis of the considered model behavior we
 present also in Fig. 1 the time behaviour of the  mean photon number of the cavity mode for $R=0.5$ and  the same value on initial field intensity.
 It is easily seen by means of calculations that collapse and revival
 time for field intensity is slightly dependent on the factor $R$. The more interesting features of the purity behaviour
 for single atom is that
 system returns most closely to a pure state of atom and field at the half-revival time, during the collapse, when mean photons number or
 population appears static \cite{Knight},\cite{Ban}. At the beginning of the collapse region $Tr (\rho_{at})^2 = 0.5$ and the atom-field state
  is maximally correlated or entangled. For two identical  two-level atoms the perfect revival of purity is absent
  but this has two local maximum
  during the collapse time.  The
   minimum value of $Tr (\rho_{at})^2$ at the beginning of the collapse region is equal approximately 0.38, whereas for
   maximally mixed two-atom system state this parameter is equal 1/4.
   This means that the system does not become completely unpolarized under the influence of the cooperative interaction.
   But for intermediate values of relative differences of two coupling constants $R$ the system behaviour is radically altered.
 The system returns most closely to a pure state at the  revival region around $t_R = \overline{n}^{1/2} \pi/g_1$. The $Tr (\rho_{at})^2$ oscillates
 in this region and reaches  maximum value which is  equal to 0.62.  At the beginning of
 the collapse region
 $Tr (\rho_{at})^2 $ is little more than 0.27 and  the maximum
 degree of entanglement between the field and atoms at this moment is  restricted.

  The more successive analysis of   atomic-field  state  as well as the   von Neumann entropy
 behaviour investigation in the considered model is the aim of a future publication.

 This work was supported by RFBR grant
 04-02-16932.


\newpage
\begin{figure}
\resizebox{90mm}{50mm}{\includegraphics{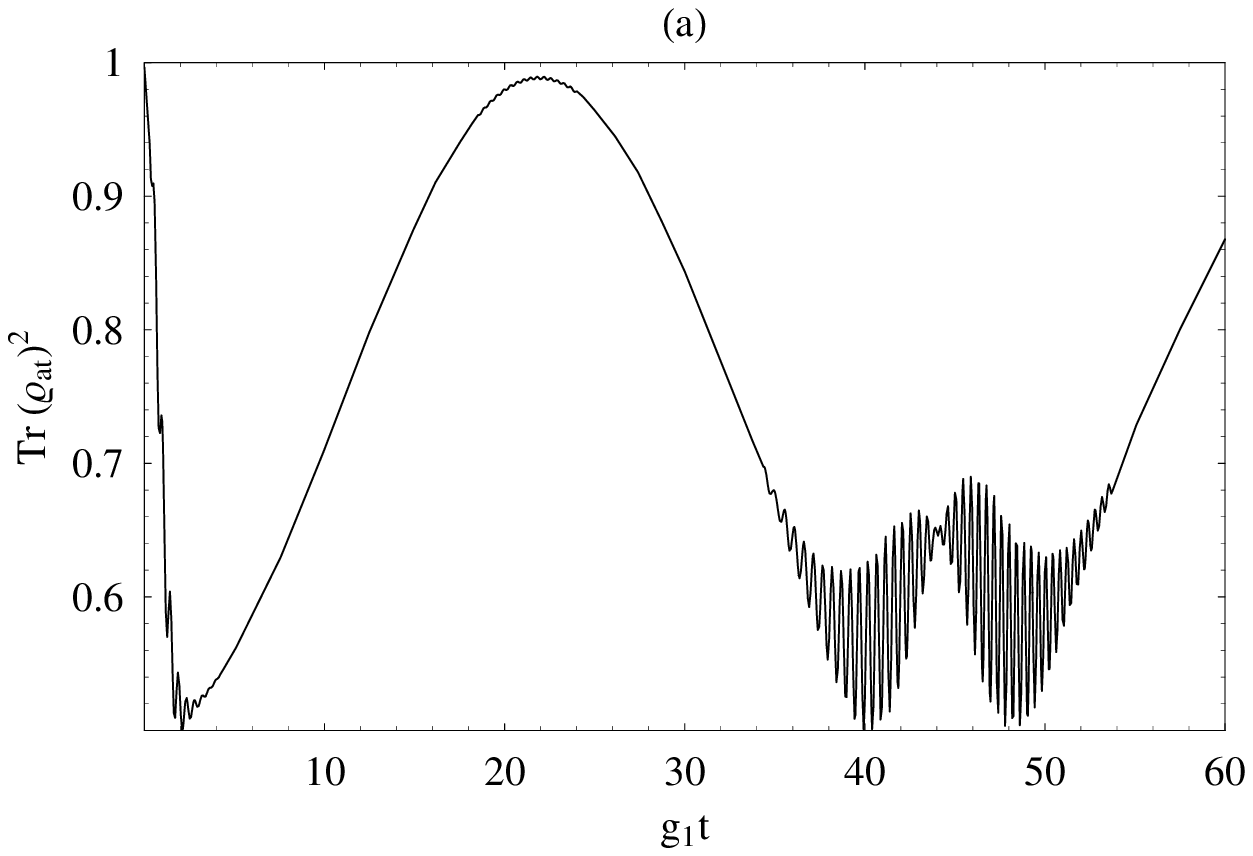}}
 \resizebox{90mm}{50mm}{\includegraphics{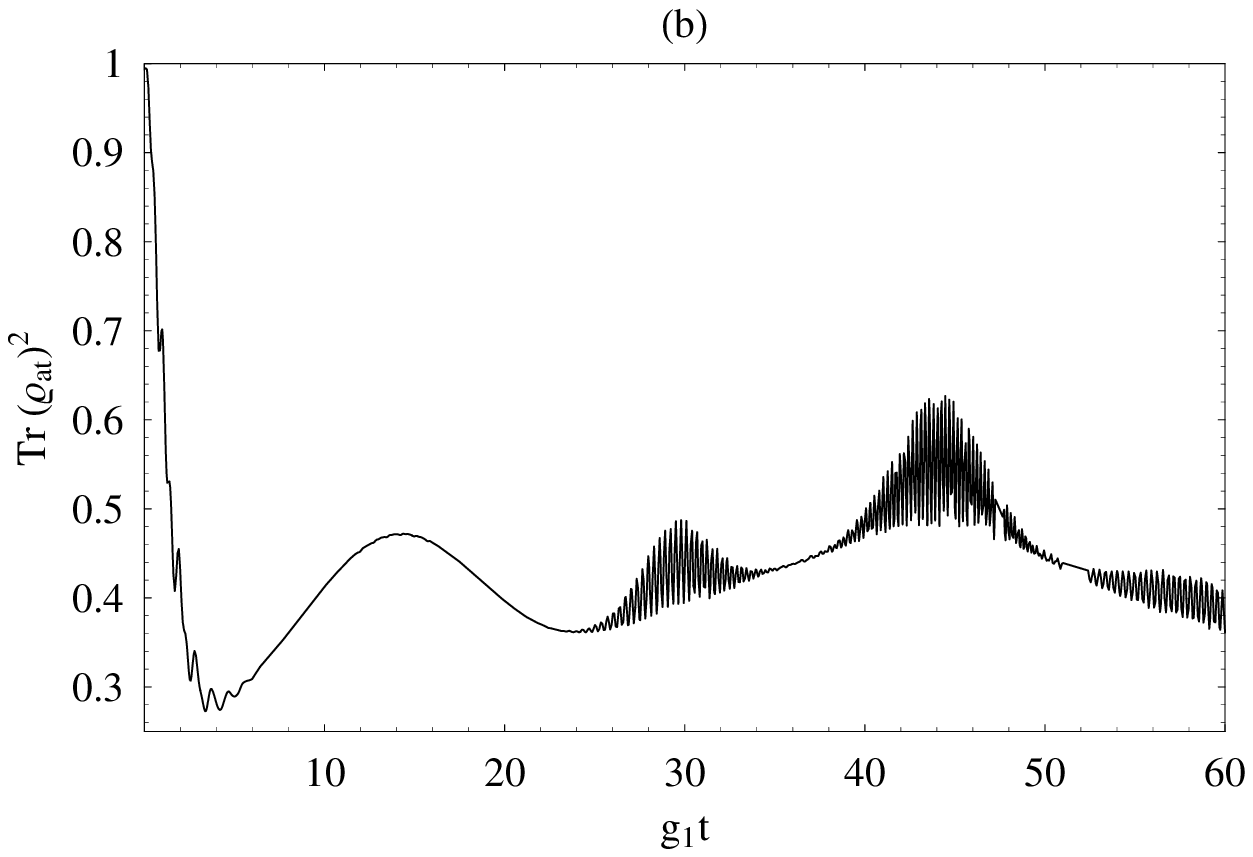}}
 \resizebox{90mm}{50mm}{\includegraphics{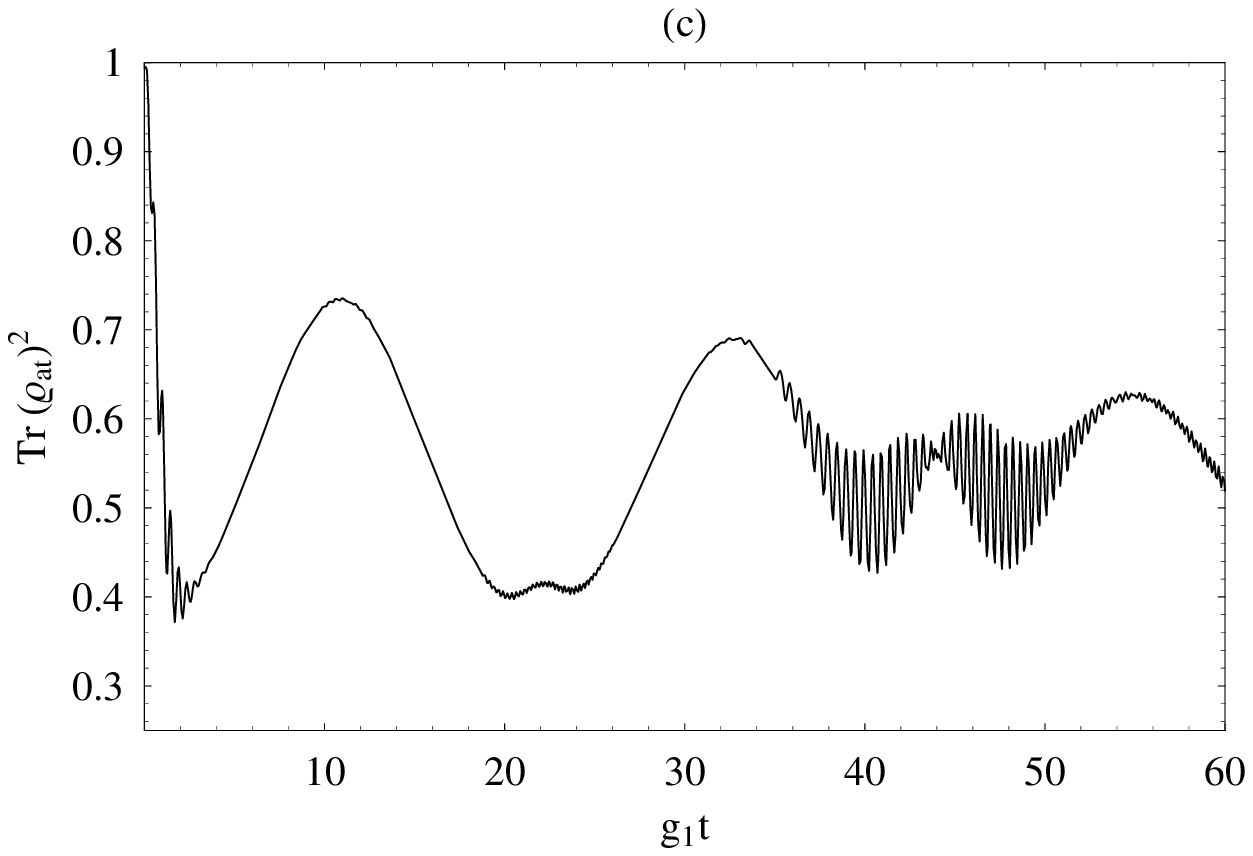}}
 \resizebox{90mm}{50mm}{\includegraphics{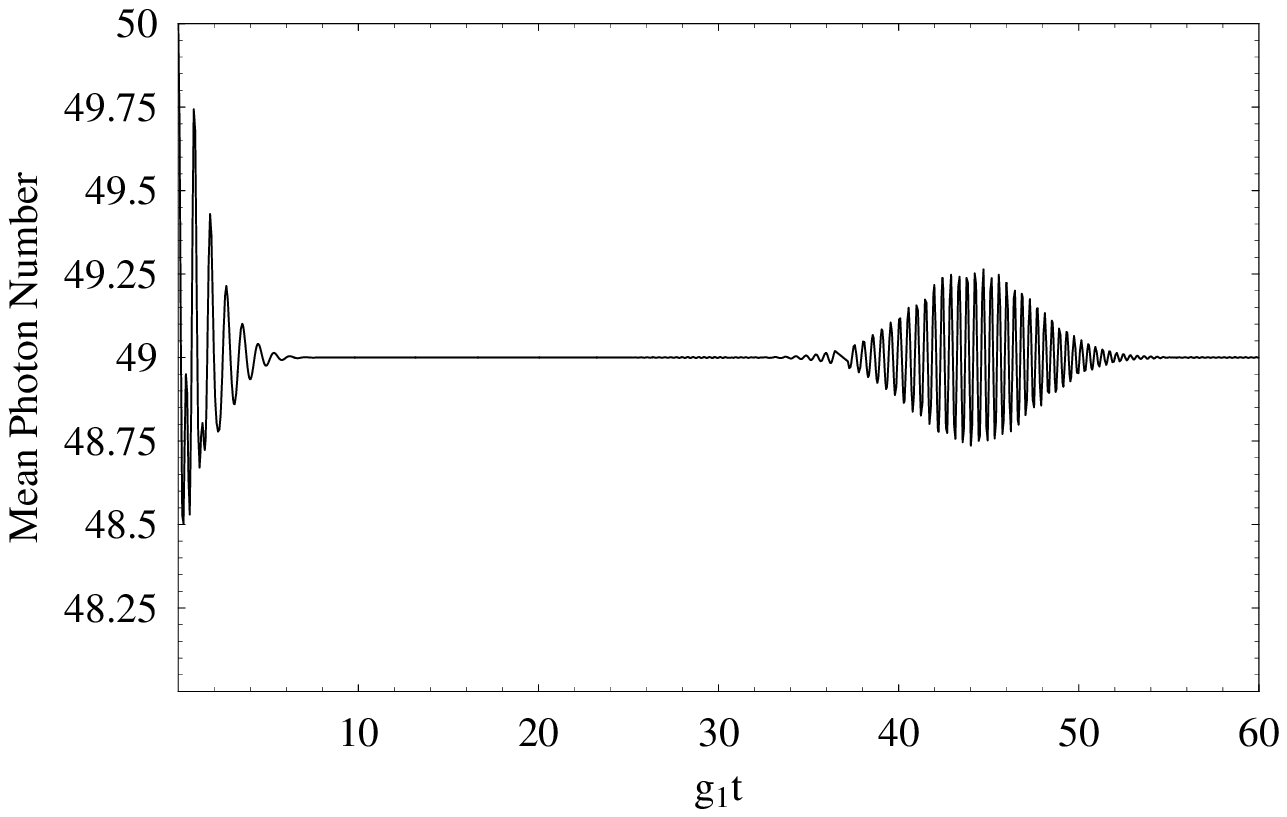}}
 \caption{Atomic purity and mean photon number against scaled time $g_1 t$ for model with $\overline{n} = 50$ and
 $R = 0$ (a), $0.5$ (b), $1$ (c) and $0.5$ (photon number).}
\end{figure}

\end{document}